\documentclass{ws-mpla}

\begin{document}

\markboth{L.Tolos, A.Ramos and T.Mizutani}
{Charm mesons at FAIR}

\catchline{}{}{}{}{}

\title{Charm mesons at FAIR}

\author{\footnotesize Laura Tol\'os$^1$, Angels Ramos$^2$ and Tetsuro Mizutani$^3$}

\address{$^1$FIAS. J.W.Goethe-Universit\"at.\\
 Ruth-Moufang-Str. 1, 60438 Frankfurt (M), Germany
\footnote{e-mail:tolos@fias.uni-frankfurt.de}\\
$^2$Dept. d'Estructura i Constituents de la Mat\`eria. Universitat de Barcelona\\
Diagonal 647, 08028 Barcelona, Spain\\
$^3$Department of Physics, Virginia Polytechnic Institute and State
University\\
Blacksburg, VA 24061, USA}

\maketitle

\pub{Received (Day Month Year)}{ Revised (Day Month Year)}

\begin{abstract}
The in-medium properties of charm mesons ($D$ and $\bar D$) in a hot and dense matter are studied.  A self-consistent coupled-channel approach is driven by a broken
SU(4) $s$-wave  Tomozawa-Weinberg interaction supplemented
by an attractive isoscalar-scalar term. As medium effects, we include Pauli blocking, baryon
mean-field bindings, and $\pi$ and open-charm meson self-energies. The dynamically generated $\tilde\Lambda_c$ and
$\tilde\Sigma_c$ resonances in the $DN$ sector 
remain close to their free space position but acquire large widths. The resultant $D$ meson
spectral function, which shows a single pronounced quasiparticle peak close to the 
free mass that
broadens with increasing density, also has a long low energy tail associated with  smeared $\tilde\Lambda_c N^{-1}$, $\tilde\Sigma_c N^{-1}$ configurations. The low-density approximation for the $\bar
D N$ is questionable already at subsaturation densities. We touch upon the implication of our study for $J/\Psi$ suppression at FAIR.
\end{abstract}

\ccode{PACS Nos.: 12.38.Lg, 14.20.Lq, 14.20.Jn, 21.65.-f }

\vspace{0.3cm}

The future CBM (Compressed Baryon Matter) experiment of the FAIR 
project at GSI will investigate, among others, the properties of open and hidden charmed  mesons in a 
hot dense baryonic environment. The  $J/\Psi$ suppression observed at higher energies might also take place at CBM conditions. 
A simple explanation for this via
$J/\Psi \to D \bar D$ dissociation assuming the in-medium meson mass
reduction of Refs.~\cite{MED}  may be inadequate due to the strong $DN$ coupled
channels. A self-consistent coupled-channel approach \cite{TOL04,TOL06,LUT06,MIZ06,TOL07} should be due
taking into account possible resonance generations.

\begin{figure}[htb]
\begin{minipage}{\textwidth}
\includegraphics[height=5.8cm,width=0.49\textwidth]{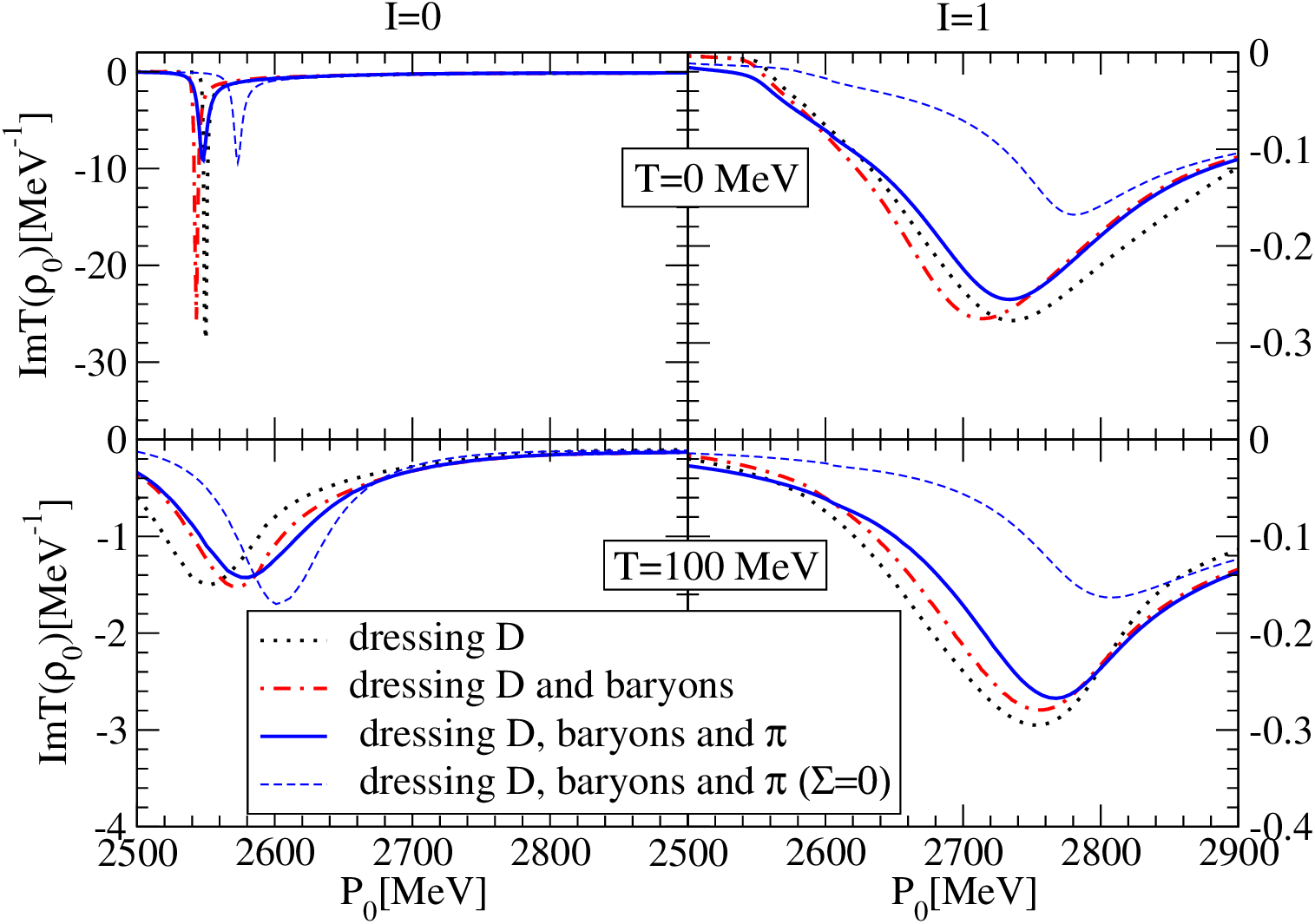}
\includegraphics[height=5.8cm,width=0.49\textwidth]{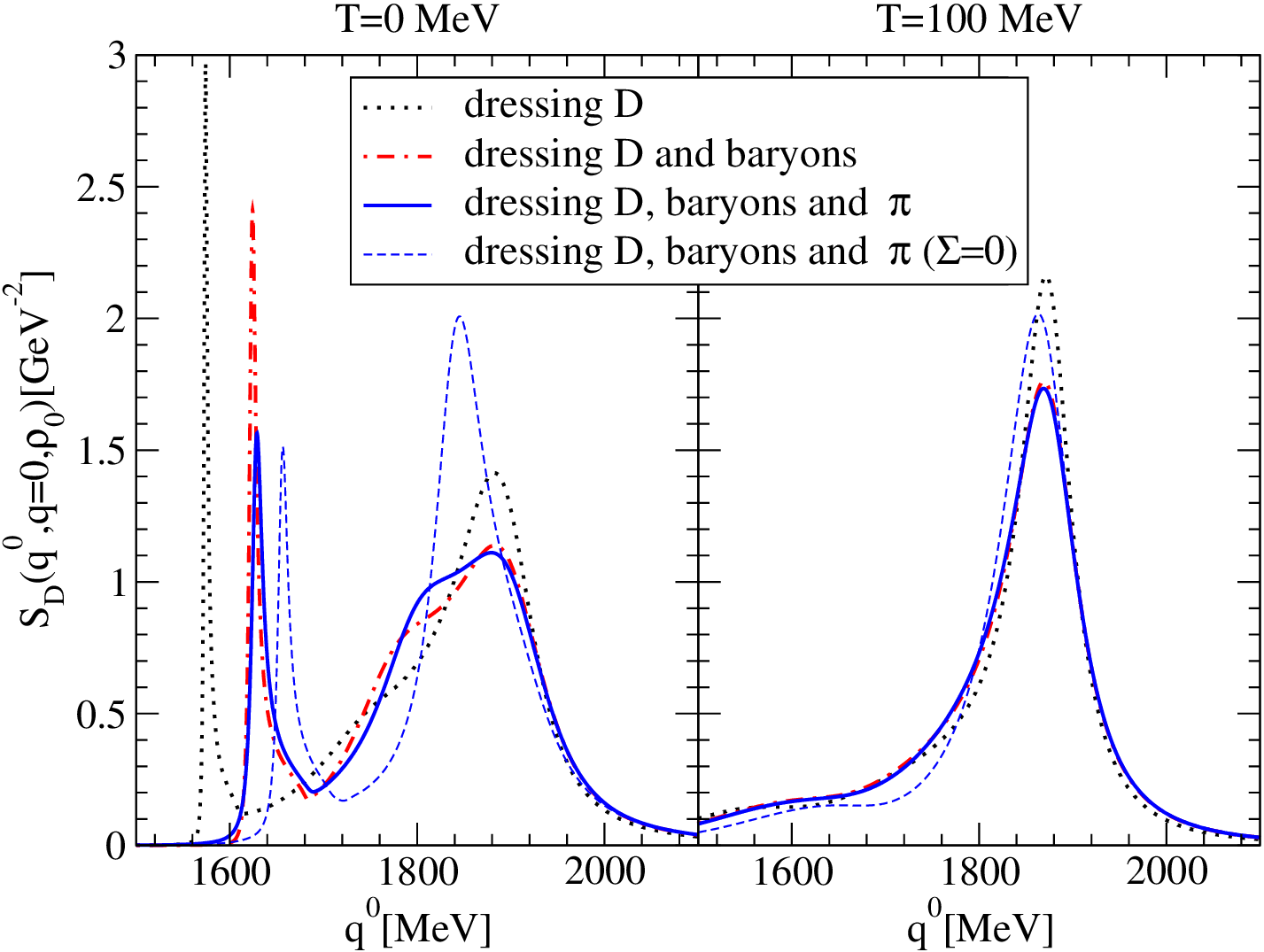}
\end{minipage}
\caption{Left: $I=0$ $\tilde\Lambda_c$ and $I=1$ $\tilde\Sigma_c$ resonances in a hot medium. Right: 
The $q=0$ $D$ meson spectral function at $\rho_0$ for $T=0, 100$ MeV }
\label{fig1}
\end{figure}

 In this work, we study the spectral
properties of $D$ and $\bar D$ mesons in nuclear matter at finite temperature
within a self-consistent coupled-channel approach \cite{TOL07} and examine the possible implications in the $J/\Psi$ suppression at FAIR.

 The $D$ and $\bar D$ self-energies at finite temperature \cite{TOL07} are obtained from a self-consistent coupled-channel calculation, whose driving term  ($V$) is a broken SU(4) $s$-wave Tomozawa-Weinberg (TW) interaction supplemented by an attractive 
 isoscalar-scalar term ($\Sigma_{DN}$). The multi-channel transition matrix ($T$), $T=V\,+\,V\,G\,T$,  is solved using a cutoff regularization \cite{MIZ06}, which is fixed by reproducing the position and the width of 
the $I=0$ $\Lambda_c(2593)$ resonance. A new $\Sigma_c$ resonance
in the $I=1$ channel is generated around 2800 MeV.

The in-medium finite temperature solution is obtained 
incorporating  Pauli blocking effects, mean-field bindings of baryons via a temperature-dependent 
$\sigma -\omega$ model, and $\pi$ and open-charm meson self-energies in the intermediate propagators. 

The $D$($\bar D$) self-energy ($\Pi_{D(\bar D)}$) is obtained self-consistently summing 
the in-medium $T_{D(\bar D)N}$ amplitudes over the thermal nucleon Fermi
distribution and the in-medium spectral function reads
\begin{equation}
S_{D(\bar D)}(q_0,{\vec q}, T)= -\frac{1}{\pi}\frac{{\rm Im}\, \Pi_{D(\bar D)}(q_0,\vec{q},T)}{\mid
q_0^2-\vec{q}\,^2-m_{D(\bar D)}^2- \Pi_{D(\bar D)}(q_0,\vec{q},T) \mid^2 } \ .
\label{eq:spec}
\end{equation}

On the l.h.s. of Fig.~\ref{fig1} we show the in-medium  $I=0$
$\Lambda_c(2593)$ and $I=1$ $\Sigma_c(2770)$ resonances, denoted as $\tilde\Lambda_c$ and $\tilde\Sigma_c$,
respectively, at saturation density ($\rho_0=0.17 \ {\rm fm^{-3}}$) for three different
cases:  i)  the self-consistent dressing of $D$ mesons only (dotted
lines), ii) including the mean-field binding of baryons (MFB) (dash-dotted lines)
and iii) with MFB and the pion self-energy dressing (PD) (solid lines). The thick lines correspond to model A ($\Sigma_{DN} \neq 0$) while
the thin-dashed lines are case (iii) within model B ($\Sigma_{DN}=0$).

At $T=0$ the medium modifications lower the 
position of the $\tilde\Lambda_c$ and $\tilde\Sigma_c$ resonances with respect
to their free  values, in particular with the inclusion of MFB. Their widths, which increase due to $\tilde
Y_c N \rightarrow \pi N \Lambda_c, \pi N \Sigma_c$  processes, differ according to the phase space available.
The PD has a small effect in the resonances because of reduced  charm-exchange channel couplings. Still it is seen in the positions and widths through the
absorption of these resonances by one and two nucleon processes when the pion self-energy is incorporated.
Moreover, models A and B show qualitatively similar
features.

The Pauli blocking effects are reduced at finite temperature 
due to the smearing of the Fermi surface. Both resonances move up in energy 
closer to their free space position while they are smoothen out, as seen in Ref.~\cite{TOL06}.  At
$T=100$ MeV, $\tilde \Lambda_c$ is  at 2579 MeV and
$\tilde \Sigma_c$ at 2767 MeV for model A, while model B generates both
resonances  at higher energies: $\tilde \Lambda_c$ at 2602 MeV and
$\tilde \Sigma_c$ at 2807 MeV.

We display in the r.h.s. of Fig.~\ref{fig1} the $D$ meson spectral function at zero
momentum for $\rho_0$ in cases (i) to (iii) for model A (thick lines) and in 
case (iii) for model B
(thin line). Two peaks
appear in the spectral function at $T=0$: that corresponding to 
$\tilde \Lambda_c N^{-1}$ excitations at lower energies and at higher energies the
quasi(D)-particle  peak  mixed with  the $\tilde \Sigma_c N^{-1}$ state. The
lower energy mode goes up by about $50$
MeV relative to (i) when MFB effects are included: the meson
requires to carry more energy to excite the $\tilde \Lambda_c$ in order to
compensate for the attraction felt by the
nucleon. The same effect is observed for the $\tilde \Sigma_c N^{-1}$
configuration that mixes with the quasiparticle peak. As expected, the PD does
not alter much the position of $\tilde \Lambda_c N^{-1}$ excitation or the
quasiparticle peak. For model B (case (iii) only), the absence of the
$\Sigma_{DN}$ term moves the $\tilde \Lambda_c N^{-1}$ excitation closer to
the  quasiparticle peak, while the latter completely mixes with the $\tilde
\Sigma_c N^{-1}$ excitation. 

Those structures are diluted when finite temperature effects are included while the quasiparticle peak gets closer to its free
value and narrower. In this case, the self-energy
receives contributions from $DN$ pairs at higher momentum where the interaction is weaker.

In the $\bar D N$ sector, we start by calculating the $\bar D N$ scattering lengths.
 For model A (B) those are $a^{I=0}=0.61 \ (0)$ fm and   $a^{I=1}=-0.26
\ (-0.29)$ fm.   The zero value of the $I=0$ scattering length for model B is
due to the vanishing coupling coefficient of the TW  $\bar
DN$ interaction. This is in contrast to the repulsive $I=0$ scattering length
in Ref.~\cite{LUT06}, while agreement is found in the $I=1$ contribution.
In the case of model A, the non-zero value of the $I=0$ scattering length is
due to the magnitude of the $\Sigma_{DN}$ term. Our results are consistent with
those of a recent calculation based on meson- and one-gluon exchanges
\cite{HAI07}. The $\bar D$ mass shift in cold nuclear matter for both A and B models  is repulsive due to the
$I=1$ dominant component. We also find that, despite the
absence of resonances in the $\bar D N$ interaction, the low-density 
$T \rho$ approximation is questionable at relatively low densities.

\begin{figure}[htb]
\begin{center}
\includegraphics[width=0.7\textwidth]{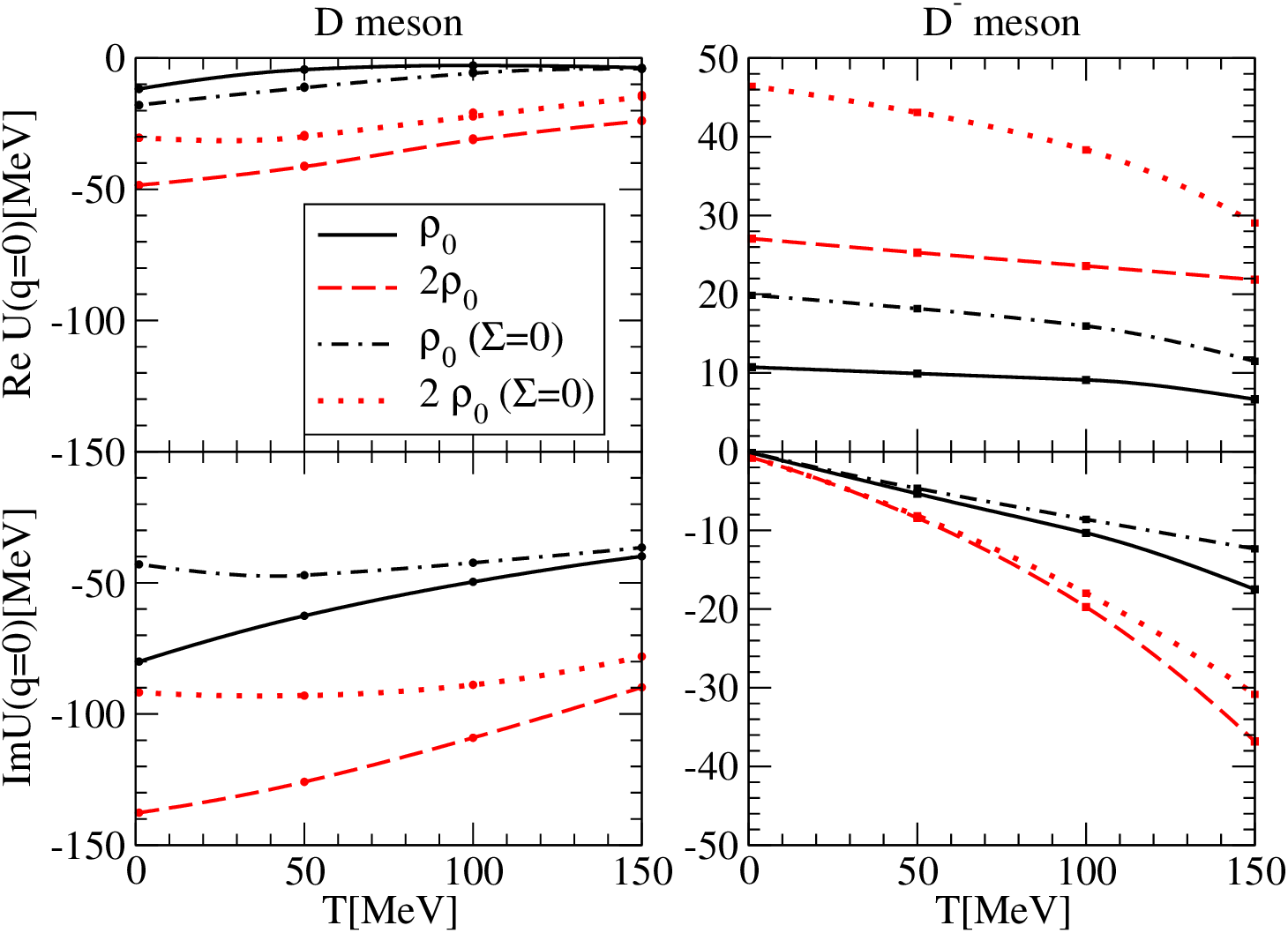}
\caption{The $D$ and $\bar D$ potentials as function of temperature } \label{fig:DDbar}
\end{center}
\end{figure}

Finally, in Fig.~\ref{fig:DDbar} we compare the $D$ and $\bar D$ optical potentials at $q=0$ MeV/c as
functions of temperature for $\rho_0$ and $2\rho_0$. For model A (B) at $T=0$ and $\rho_0$, the
$D$ meson  feels an attractive potential of $-12$ ($-18$) MeV  while the $\bar
D$ feels a repulsion of $11$ ($20$) MeV. A similar shift for $D$
meson mass is obtained in Ref.~\cite{TOL06}. The temperature dependence of the
repulsive real part of the ${\bar D}$ optical potential is very weak, while the
imaginary part increases steadily due to the increase in the collisional width. The
picture is somewhat different for the $D$ meson due to the overlap of the
quasiparticle peak with the $\tilde{\Sigma}_c N^{-1}$ mode. Therefore, the in-medium behavior of the  $\tilde{\Sigma}_c N^{-1}$ mode is determinant for
understanding the effect of the $\Sigma_{DN}$ term on the $D$ meson potential.

Regarding the $J/\Psi$ suppression in an hadronic environment, the in-medium $\bar D$ mass increases about $10-20$ MeV while the tail of the quasiparticle peak of
the $D$ spectral function extends to lower "mass" due to the thermally spread
$\tilde Y_c N^{-1}$. However, it is very unlikely that this lower tail
extends as far down by 600 MeV with sufficient strength for the $J/\Psi \rightarrow D \bar D $ to go.  So the only way for the $J/\Psi$ suppression to take
place is by cutting its supply from the excited charmonia ($\chi_{c\ell}(1P)$
or $\Psi'$) due to multi-nucleon absorption.



\end{document}